# Classical to quantum transition of heat transfer between two clusters

Shiyun Xiong[1,2], Kaike Yang[3], Yuriy A. Kosevich[1,2,3], Yann Chalopin[1,2], Roberto D'Agosta[3,5], Pietro Cortona[6], Sebastian Volz[1,2]

[1] CNRS, UPR 288 Laboratoire d'Energétique Moléculaire et Macroscopique, Combustion (EM2C), Grande Voie des Vignes, 92295 Châtenay-Malabry, France
[2] Ecole Centrale Paris, Grande Voie des Vignes, 92295 Châtenay-Malabry, France
[3] Nano-Bio Spectroscopy Group and ETSF Scientific Development Centre Departamento Fisica de Materiales, Universidad del Pais Vasco UPV/EHU, E-20018 San Sebastian, Spain
[4] Semenov Institute of Chemical Physics, Russian Academy of Sciences, 119991 Moscow, Russia
[5] IKERBASQUE, Basque Foundation for Science, E-48011, Bilbao, Spain.
[6] Laboratoire Structures, Propriétés et Modélisation des Solides, UMR CNRS 8580, École Centrale Paris, 92295, Châtenay-Malabry, France

**Abstract:** Heat transfer between two silica clusters is investigated by using the non-equilibrium Green's function method. In the gap range between 4 Å and three times the cluster size, the thermal conductance decreases as predicted by the surface charge-charge interaction. Above five times the cluster size, the volume dipole-dipole interaction predominates. Finally, when the distance becomes smaller than 4 Å, a quantum interaction where the electrons of both clusters are shared takes place. This quantum interaction leads to the dramatic increase of the thermal coupling between neighbor clusters due to strong interactions. This study finally provides a description of the transition between radiation and heat conduction in gaps smaller than a few nanometers.

With the recent developments of nanotechnology, electronic devices continue to scale down in dimension and scale up in power density [1]. As a result, near-field radiation starts to play a notable role in the thermal design at nanoscales. Recently, it has been demonstrated both theoretically [2-7] and experimentally [8-10] that heat transfer through near-field radiation between two parallel plates or between a sphere and a plane can be several orders of magnitude larger than the black body limit over a limited range of frequency. This clearly corroborates the fact that when the gap between two objects is smaller than the characteristic photon wavelength, a different physical behavior emerges in which near-field radiation and acoustic phonon tunneling significantly contribute to heat transfer [11,12].

The first measurement of the radiative heat flux between two dielectric materials separated by a nanoscale gap distance has recently been performed by Narayanaswamy, Shen and Chen [8,9]. Their results led them to conclude that the proximity-force approximation is not valid for near-field radiation heat transfer. Shortly after, Rousseau et al. [10] also measured the heat transfer in the near-field regime. Interestingly, and in contrast with Narayanaswamy's conclusions, these later results confirmed the proximity approximation. The difficulty in performing such experiments makes it probable that heat transfer at the nanoscale will continue to be debated, as commented by Kittel [13]. Near-field radiation under the dipole/multipole approximation has been extensively investigated on a theoretical basis. Nonetheless, mechanisms taking place for separation distances shorter than 10 nm remain unclear. This range of separation distances may not be directly accessible by experiments due to the difficulty in fabricating well-defined planes and spheres at those scales. At the same time, as modern nanostructures might be smaller than 10 nm and are separated in some cases by only a few fractions of a nanometer, this range of lengths is of great interest to those who design nanoscale devices [13,14]. From a fundamental point of view, this domain is also involving the less understood transition from a classical charge-charge interaction, logically described as a radiation in the near-field, to a chemical bond interaction, yielding pure heat conduction. By means of molecular dynamics simulations, Domingues et al [15] found a transition regime characterized by a thermal conductance larger than the contact conductance. But the largest value exceeded the upper physical limit. Using ultra-high vacuum inelastic scanning tunneling microscopy, a previously unknown mechanism of thermal transport- a field-induced phonon tunneling- has been reported by Altfeder et al [16]. The thermal energy transmitted through atomically narrow vacuum gap exceeds, by ten orders of magnitude, the one of blackbody thermal radiation. In fact, before these experimental findings, Kosevich [11] and Prunnila et al [12] have modeled how acoustic phonons can directly tunnel through vacuum by introducing coupling mechanisms, and both of them have shown that acoustic phonons can travel through the vacuum gap with unitary transmission and thus can lead to significant thermal conductance and heat flux.

In this letter, we estimate the heat transfer through a chain composed by identical non-contacting silica clusters by means of the phonon non-equilibrium Green's functions. We show that there are two critical vacuum gaps of about 4 Å and three to five times the cluster size. The first critical gap, of 4 Å, corresponds to a transition between the classical and the quantum regimes with strong interaction. Above this critical gap, the conductance decreases first according to a $d^{-3}$ power law, d being the distance between the center of masses, and then gradually follows a $d^{-6}$ power law, when the gap is larger than five times the cluster size. These power laws can be explained by classical surface charge-charge and volume dipole-dipole interactions, respectively. Below 4 Å, the conductance shows a much stronger dependence on gap thickness. The first critical gap is confirmed by *ab-initio* calculations showing that the electronic wave functions indeed merge when the gap becomes shorter than 4 Å.

We consider a system of identical silica clusters separated by a distance $d$ between the cluster centers and a gap $l$ (Figure 1, top). Clusters are coupled through the van Beest, Kramer, and van Santen (BKS) potential [17], composed of Coulomb and Buckingham potentials. The BKS potential provides the full physical picture of the long range electromagnetic and the short-range repulsive-attractive interactions. We consider one cluster as the reference system and the clusters on its left and right sides serve as reservoirs. The system period is illustrated in Figure 1 (top). The gap conductance $\sigma$ between two clusters is derived from the energy transmission $T_r$ as follows [18-20]:

$$\sigma = \int_0^{\omega_{max}} T_r(\omega) \frac{\partial}{\partial T}\left(\frac{1}{e^{\hbar\omega/k_B T}-1}\right)\hbar\omega \frac{d\omega}{2\pi} \quad (1)$$

where $\omega$ and $\omega_{max}$ are the energy and the Debye frequencies. $T$ refers to the mean temperature of the system, $k_B$ and $\hbar$ represent the Boltzmann and the reduced Planck's constants, respectively.

The transmission $T_r$ is obtained from a non-equilibrium Green's function approach [18-20] as $T_r(w) = Tr[\Gamma_L G_s \Gamma_R G_s^+]$. The advanced and retarded Green functions $G_s^+$ and $G_s$ can be deduced from:

$$G_s = [(w+i\text{D})^2 I - K_{ss} - \text{S}_L - \text{S}_R]^{-1}, \quad (2)$$

where $\eta$ is an infinitesimal imaginary part that maintains the causality of the Green's function and $\text{S}_L = K_{ab} g_L K_{ab}^\square$, $\text{S}_R = K_{ab} g_R K_{ab}^\square$ are the self-energies of the left and right leads, the "+" exponent indicating the Hermitian conjugation. Finally, $g_L$ and $g_R$ refer to the Green's functions of the left and right leads, $K_{ss}$ and $K_{ab}$ being the force constant matrices derived from the BKS potential, for one cluster and between neighboring clusters, respectively. The expression of the transmission also includes $\Gamma_L = i(\text{S}_L - \text{S}_L^+)$ and $\Gamma_R = i(\text{S}_R - \text{S}_R^+)$.

The thermal conductance between clusters obtained from Eq. (1) is reported in Figure 2. The conductance decreases very quickly with distance in the short gap range. The power law in this range is estimated to be about $d^{-12}$ and the absolute value of the power slightly increases with the increase of the particle size. The thermal conductance per unit cross-section indeed increases with cross section as the number of interacting pairs per atom increases. This latter number becomes larger at short distances and leads to a slight growth of the absolute exchanged power. This growth should however saturate to a maximum value as the number of interacting pairs per atom also saturates, but this limit remains beyond the maximum size under consideration here.

In the intermediate distance range, the conductance decrease with distance turns to be smoother and follows the power law $d^{-3}$, which is expected in the framework of the non-piezoelectric interactions [11]. Silicon and oxygen atoms form a dipole as shown in the inset of Figure 2 and each particle can be regarded as one macroscopic dipole, with bound charges of opposite sign at front and rear surfaces. When the distance between two clusters is comparable with the cluster size, the force per unit surface area between surface charges is proportional to $S_2/d^2$, where $S_2$ is surface area of cluster 2. According to our model, the transmission of acoustic phonons through the vacuum gap can be written as [11]:

$$|T_{aph}|^2 = \frac{1}{1+(\omega/\omega_0)^2} \tag{3}$$

where represents the effective width of the acoustic phonon pass band through the gap; is proportional to the modulus of the derivative of the force per unit surface area with respect to the gap width, and as a result, $\omega_0 \propto S_2/d^3$. The total thermal conductance is given by the integral of transmission (3) over all frequencies times surface area of cluster 1 $S_1$ and is proportional to $S_1\omega_0$, and hence is characterized by the scaling $S_1S_2/d^3$ [11]. This means that the conductance in this range of distance $d$ is performed mainly by acoustic phonons, which is in agreement with our results obtained from Green's function (shown below). Interestingly, the slope transition in the log scale occurs at the same gap distance $l$=4 Å whatever the cluster diameter. When the gap increases further, i.e. the distance between two neighboring clusters becomes much larger than the particle size, the energy transfer between two clusters is performed by optical phonon exchange through dipole-dipole interaction [15], following the Foerster energy transfer with a $1/d^6$ decay law [21]. The transition from the charge-charge to dipole-dipole interaction occurs smoothly when $d$ is around three to five times the cluster size. Furthermore, in the charge-charge interaction region, the conductance at a given gap width $l$ follows a $D^{3.85}$ scaling law, while in the dipole-dipole interaction range, the conductance varies with diameter according to $D^{6.5}$ for a given distance $d$. These findings further confirm our proposed mechanism of surface charge-charge and volume dipole-dipole heat transfer since the total conductance is proportional to the product of clusters surface areas $S_1S_2$, that is to $D_1^2 D_2^2/d^3$, for surface charge-charge interaction, while the conductance is proportional to the product of clusters volumes $V_1V_2$, that is to $D_1^3 D_2^3/d^6$, for volume dipole-dipole interaction.

To validate our predictions, molecular dynamics (MD) simulation results as taken from reference [14] are plotted in Figure 2 for comparison. A clear agreement between MD and Green's function predictions appears in the long distance range. But there is no intermediate region in MD predictions and the conductance from Green's function is several orders of magnitude smaller than the one yielded from MD in the small gap range. Also in contrast to MD simulations, no conductance decrease is found right before the contact in our Green's function calculations. Instead, the conductance increases monotonically while the gap decreases. In fact, the maximum conductance before contact predicted by MD simulations exceeds the physical upper-limit $\sigma_{max}$ as shown in Figure 2. This limit is calculated from the maximum energy $3Nk_B(T_1-T_2)$ possibly transferred between two neighbor clusters of $N$ atoms each, set to temperatures $T_1$ and $T_2$. Considering the fastest transfer characterized by the highest mode frequency $f_{max}$, the maximum conductance is obtained as $\sigma_{max} = 3Nk_B f_{max}$. The MD predicted conductance just before the contact is one or two orders of magnitude larger than the maximum value while the non-equilibrium Green's function predictions give estimations below this limit.

To understand the origin of the change in the dependence of the conductance to the distance $d$, we performed *ab-initio* calculations (ABINIT code [22]) of the electron densities for two silica planes schematically shown in the below panel of Figure 1, and separated by vacuum gaps ranging from 0 to 6 Å. As each plane consists in a 1x1x2 supercell, the two cells axis being perpendicular to the interacting surfaces. Each unit cell contains twelve atoms and the simulation box includes four cells and 48 atoms. Experimental data for the atomic positions are used and the exchange-correlation

Hamiltonian is treated within the generalized gradient approximation with the Perdew-Burke-Ernzerhof functional [23]. Fritz-Haber Institute pseudopotentials [24] are adopted for Si and O atoms. The cut-off energy is set to 820 eV and the k-grid size to 4x4x1.

As revealed by Figure 3, the electron density is nonzero in the middle of the gap, when the gap is smaller than 4 Å but decreases rapidly as the gap widens from 0 to 4 Å. The electron density reaches zero in the middle of the gap when $l$ increases beyond 4 Å and the zero electron density domain extends when further increasing the gap. This indicates that the electron wave functions of both sides actually overlap in the short gap range when $l < 4$ Å to form a bond. In this region, the atoms of both sides are connected through a single electronic wave function instead of interacting through electromagnetic forces relating two separated wave functions. Beyond 4 Å, near field radiative heat transfer can be described by Maxwell equations while the quantum Schrödinger equation has to be considered when $l < 4$ Å. Since the bonds between atoms in silica are covalent, we may call the bond before the contact as 'pseudo-covalent'. With the formation of those latter bonds, the force between two neighbors dramatically increases beyond the force produced by electromagnetic waves. As a result, heat transfer shifts from radiative to conductive, also leading to a slope change of the thermal conductance in the small gap range.

To check the relative contribution of acoustic and optical phonons to heat conduction, we now turn to our previous modeling of the transmission of acoustic phonon modes through a vacuum gap as shown in equation (3). Since $\omega_0$ represents the effective width of the acoustic phonon pass-band through the gap and it is proportional to the derivative of the force between clusters with respect to the gap width, it decreases with the increase of the gap width. Consequently, the acoustic phonon cut-off frequency decreases when the gap widens and the frequency range of allowed transmission converges to zero.

Figure 4 reports the cumulative transmission coefficient from one cluster to its neighbor as a function of frequency and distance $l$. The cumulative transmission function increases continuously for the smallest gap of 1 Å (black line) reflecting a continuous dependence of the transmission on frequency. The continuous decrease of the transmission as the frequency $\omega_0$ reduces to zero reveals that the modes involved are indeed acoustical ones.

When the gap width is slightly increased from 1 Å to 4 Å, the cumulative transmission function dramatically decreases and includes both a continuum at low frequencies and a set of jumps due to a discrete transmission at higher frequencies as highlighted by the inset of Figure 4. In qualitative agreement with the model of Eq. (3), widening the gap indeed results in a decrease of the acoustic frequency pass band, which uncovers the presence of optical contributions appearing as peaks in the transmission spectrum. A careful analysis of our data shows that the frequency range of the acoustic phonons continuum reduces to zero as the gap width reaches five times of particles size and accomplishes most of the heat transfer when the gap width is smaller than three times the particle size. The discrete set of modes also progressively disappear when the gap width is further increased and only the modes related to force constants of long range interactions remain when the gap is enlarged and those also gradually disappear as those long range interactions vanish.

By considering the phonon-induced interactions of the gap edges, Kosevich [11] and Prunnila and Meltaus [12] have shown independently that acoustic phonons could transmit energy between separated bodies by tunneling through vacuum gap, which can led to a significant thermal conductance enhancement and which is consistent with our findings. Accordingly, Altfeder et al. [16] observed phonon tunneling from a sharp STM tip into a gold film at a vacuum gap distance of 3 Å. The authors claim that the tunneling effect is driven by surface electron-acoustic-phonon interaction. This result supports our argument stating that acoustic phonons are predominant in the phonon tunneling through small gaps.

In conclusion, the non-equilibrium Green's function technique has been implemented for calculating the heat transfer between two silica clusters. We found that the studied gap range can be divided into three parts with two critical gaps of 4 Å and three to five times the cluster size. The heat transfer regimes are characterized by the decay power laws of $d^{-12}$, $d^{-3}$ and $d^{-6}$, successively. The critical gap of 4 Å corresponds to the classical to quantum transition beyond which heat transfer between neighboring clusters follows the classical law prescribed by surface charge-charge (intermediate range) and volume dipole-dipole (long-range) interactions, while the heat flux drastically increases for the gap distance below 4 Å. Near-field radiation clearly captures the thermal interaction above $l$=4 Å, but the heat transfer below this distance is dominated by heat conduction as we have shown that electrons are actually forming a chemical bond in the gap. Our results thus provide a deeper insight into understanding the behavior of the transition between radiation and heat conduction in gaps smaller than a few nanometers.

K. Y. and R.D'A. acknowledge the financial support from CONSOLIDER INGENIO 2010: NANOTherm (CSD2010-00044), Diputacion Foral de Gipuzkoa (Q4818001B), the European Research Coun- cil Advanced Grant DYNamo (ERC-2010-AdG-267374), Spanish Grants (FIS2010-21282-C02-01), Grupos Con- solidados UPV/EHU del Gobierno Vasco (IT578-13), Ikerbasque, and MAT2012-33483.

# Figure Captions

**Figure 1**. (Color online) Schematics of the silica cluster systems considered in the Green's function calculations (top) and the *ab-initio* computations (bottom). For Green's function calculation, the clusters are *N*x*N*x*N* unit cells cubes with $SiO_2$ lattice constant of 4.52 Å. In *ab-initio* calculations, two parallel Silica planes separated with different gap distances are used and electron densities inside the gaps are calculated with this model.

**Figure 2.** (Color online) Thermal conductance between two neighboring clusters at 300K for different cluster sizes versus the distance *d* indicated in Figure 1. In our calculations, the cluster is a cube *N*x*N*x*N* unit cells in volume. The diameter *D* is set in such a way that the sphere volume is equivalent to that of the simulated cube. The distance *d* was used as the abscissa instead of the gap distance *l* in order to discriminate the curves otherwise superimposed. The MD results are taken from reference [14] where the same BKS potential parameters as those adopted in this work were used.

**Figure 3.** (Color online) Ab-initio computation of the electron density generated by two parallel silica films separated by different gap widths.

**Figure 4**. (Color online) Angular frequency dependent cumulative phonon transmission for different gap distance *l* in the cluster of diameter D = 1.3 nm. Inset: phonon transmission function versus angular frequency at low frequencies.

# Figures

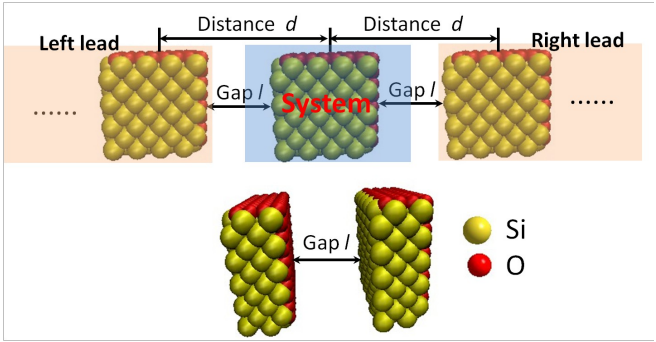

Figure 1

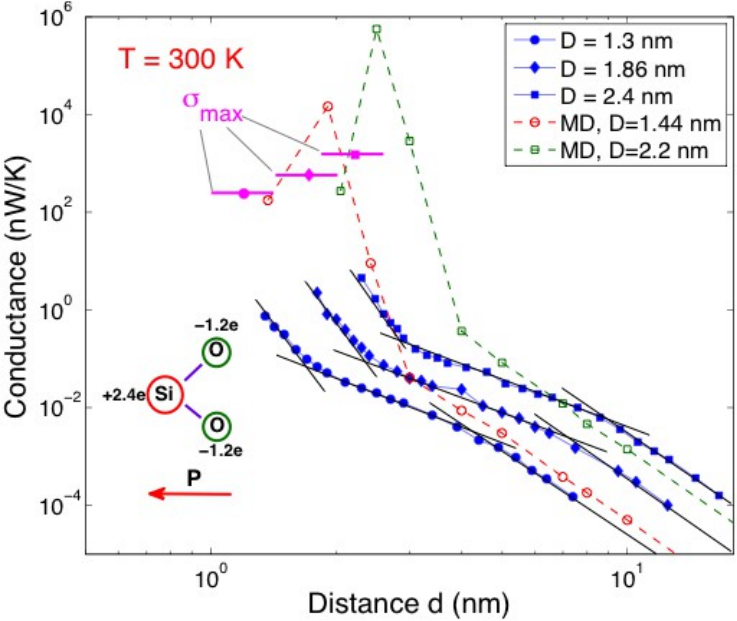

Figure 2

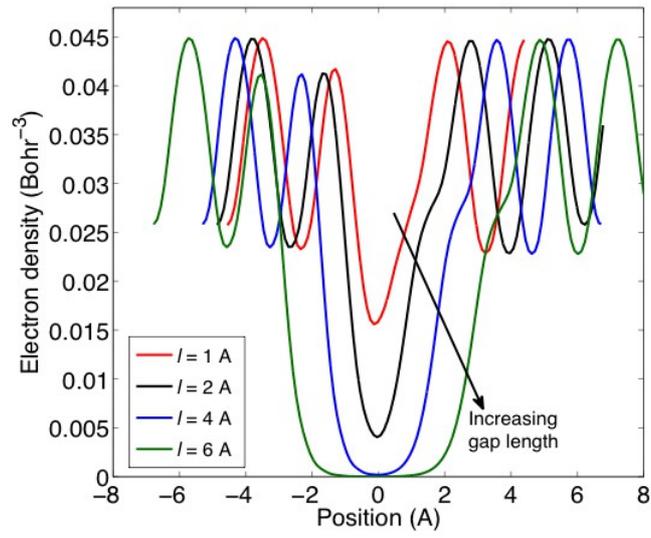

Figure 3

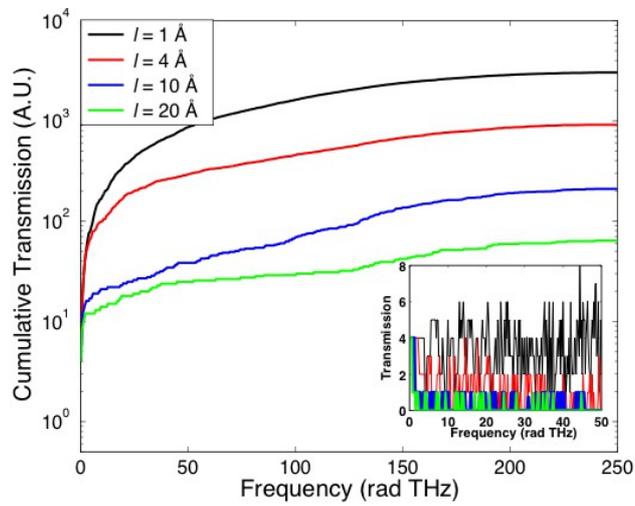

Figure 4